\documentclass[twocolumn,aps,prl,showpacs,amsmath]{revtex4}
\usepackage{graphicx}
\begin{document}


\title{ Evidence of a spin resonance mode in the iron-based superconductor Ba$_{0.6}$K$_{0.4}$Fe$_{2}$As$_{2}$ from scanning tunneling spectroscopy }

\author{Lei Shan$^1$}\email{lshan@aphy.iphy.ac.cn}
\author{Jing Gong$^1$,Yong-Lei Wang$^1$, Bing Shen$^1$, Xingyuan Hou$^1$, Cong Ren$^1$, Chunhong Li$^1$}
\author{Huan Yang$^{1,2}$, Hai-Hu Wen$^{1,2}$, Shiliang Li$^1$}
\author{Pengcheng Dai$^{1,3}$}

\affiliation{$^1$National Laboratory for Superconductivity, Beijing National Laboratory for Condensed Matter Physics, Institute of Physics, Chinese Academy of Sciences, Beijing 100190, China}

\affiliation{$^2$National Laboratory of Solid State Microstructures and Department of Physics, Nanjing University, Nanjing 210093, China}

\affiliation{$^3$Department of Physics and Astronomy, The University of Tennessee, Knoxville, Tennessee 37996-1200, USA }

\date{\today}

\begin{abstract}
We used high-resolution scanning tunneling spectroscopy to study the hole-doped iron pnictide superconductor Ba$_{0.6}$K$_{0.4}$Fe$_{2}$As$_{2}$ ($T_c=38$ K). Features of a bosonic excitation (mode) are observed in the measured quasiparticle density of states. The bosonic features are intimately associated with the superconducting order parameter and have a mode energy of $\sim$14 meV, similar to the spin resonance measured by inelastic neutron scattering. These results indicate a strong electron-spin excitation coupling in iron pictnide superconductors, similar to that in high-$T_c$ copper oxide superconductors.

\end{abstract}

\pacs{74.55.+v, 74.70.Xa, 74.20.Mn}

\maketitle

In conventional superconductors, the electron-phonon interaction responsible for electron pairing and superconductivity was unequivocally established by tunneling and neutron scattering experiments, where dips in the second derivative of the tunneling current $d^2I/dV^2$ correspond to phonon modes observed by inelastic neutron scattering experiments \cite{EliashbergGM1960,SchriefferJR1963,McMillanWL1965}. To obtain the equivalent information for high-transition temperature (high-$T_c$) superconductors, it is important to identify the electron-boson coupling \cite{MaierTA2008} and its connections to superconductivity. For bosonic ``pairing glue'' mediated superconductors, the ``glue'' may arise from the usual electron-phonon interactions \cite{BardeenJ1957} or the exchange of particle-hole spin fluctuations characterized by the imaginary part of the dynamic susceptibility, $\chi^{\prime\prime}(Q,\omega)$, which is seen in inelastic magnetic neutron scattering measurements \cite{MonthouxP2007,MoriyaT2003RPP,EschrigM2006AIP,CarbotteJP2011RPP}. For electron-doped high-$T_c$ copper oxide superconductors, tunneling experiments using scanning tunneling microscopy on Pr$_{0.88}$LaCe$_{0.12}$CuO$_4$ have identified a bosonic excitation at an energy (10.5 meV) consistent with the neutron spin resonance \cite{NiestemskiFC2007,WilsonSD2006,ZhaoJ2011}, thus providing evidence that spin fluctuations mediate superconductivity \cite{JinK2011}. The spin resonance mode has also been observed in tunneling experiments of hole-doped copper oxide superconductors \cite{ZasadzinskiJF2001PRL,ZasadzinskiJF2006PRL,AhmadiO2011PRL}, and quantitative analysis suggested that quasiparticle interactions with this mode are important for superconductivity. In the case of iron-based superconductors \cite{HosonoH2008JACS,ChenXH2008Nature,WenHH2008EPL,ChenGF2008PRL,RenZA2008CPL,RotterM2008PRL}, spin fluctuations have been suggested to be responsible for superconductivity through sign reversed quasiparticle excitations between the isotropic hole pockets near $\Gamma$ and electron pockets near $M$ \cite{MazinII2010}(the so-called $s_\pm$-wave). Although the discovery of a neutron spin resonance in optimally hole-doped BaFe$_2$As$_2$ \cite{ChristiansonA2008,ZhangCL2011} and a bosonic structure in scanning tunneling spectroscopy (STS) of SmFeAsO$_{1-x}$F$_x$ \cite{FasanoY2010} are consistent with spin fluctuation mediated $s_\pm$-wave superconductivity, these measurements were carried out on different classes of materials. Therefore, there is no direct experimental evidence connecting neutron spin resonance to the bosonic features in STS of the same material.

In this Letter, we report high-resolution STS studies of the optimally hole-doped iron-based high-$T_c$ superconductor Ba$_{0.6}$K$_{0.4}$Fe$_{2}$As$_{2}$ ($T_c=38$ K). We found that the second derivative of the tunneling current $d^2I/dV^2$ shows clear high-bias ($eV>\Delta$) features at energies of $\sim$ 14 meV, in good agreement with the neutron spin resonance mode \cite{ChristiansonA2008,ZhangCL2011}. The ratio of the mode energy and superconducting gap suggests that the resonance mode is a magnetic exciton as observed in copper oxide superconductors, indicating that spin fluctuations are important for high-$T_c$ superconductivity.

\begin{figure}[]
\includegraphics[scale=1.05]{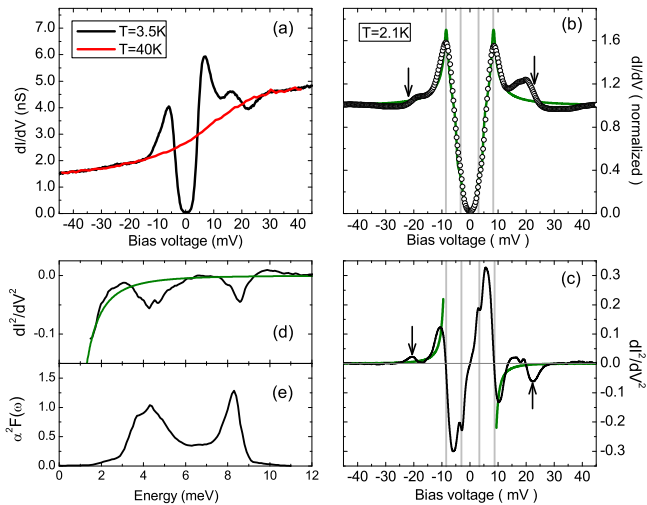}
\caption{\label{fig:fig1} (Color online) (a) Typical tunneling spectra measured below (black) and above (red) $T_c$. (b) Normalized spectrum averaged over an area of 16.7 nm $\times$ 8.4 nm. The features of two superconducting gaps are indicated by gray vertical lines. Black arrows indicate the hump-dip structures mentioned in the text, and the green line is a fit to two-gap weak-coupling BCS theory. Spectra were taken with a fixed junction resistance of 0.25 G$\Omega$. (c) Second derivative of tunneling current calculated from the data presented in (b). The green line is a BCS calculation. (d) The black curve is the relationship between $d^{2}I/dV^{2}$ vs. ($eV-\Delta$) obtained on a conventional strong-coupling superconductor Pb in an energy window outside the superconducting gap $\Delta$ \cite{McMillanWL1965}. The green line is a BCS curve. (e) Electron-phonon spectral function of Pb \cite{McMillanWL1965}. }
\end{figure}

The nearly optimally-doped Ba$_{0.6}$K$_{0.4}$Fe$_{2}$As$_{2}$ (T$_c$ = 38 K) single crystals studied here were grown with the self-flux method \cite{LuoHQ2008SST}. The spatially-resolved tunneling experiments were carried out on a low-temperature scanning tunneling microscope that was constructed in-house \cite{ShanL2011,ShanL2011prb}. A single crystalline sample was cold-cleaved \emph{in situ} and then immediately inserted into the microscope, which had been maintained at the desired temperature.

Figure~\ref{fig:fig1}(a) shows typical tunneling spectra ($dI/dV$ vs. $V$) at temperatures below and above $T_c$. At $T=3.5$ K, there were two sharp coherence peaks accompanied by zero conductance around zero bias voltage, indicating a superconducting gap. Upon warming up to $T=40$ K, the coherence peaks disappeared and the spectrum became featureless. More features can be seen in the spectrum shown in Fig.~\ref{fig:fig1}(b), which is an average over a 16.7 nm $\times$ 8.4 nm region measured at 2.1 K. As indicated by the gray vertical lines, in addition to a larger gap of 8.4 meV, a smaller gap of 3.5 meV appears as kinks on the spectrum. The two superconducting gaps are also distinguished in the second derivative of the tunneling current as shown in Fig.~\ref{fig:fig1}(c). This two-gap structure can be fitted very well to the BCS theory by combining the spectral contributions from the two superconducting gaps as shown in Fig.~\ref{fig:fig1}(b), in good agreement with our previous measurements \cite{ShanL2011,ShanL2011prb}. In general, STS detects the integrated quasiparticle density of states (DOS) over all Fermi surfaces with a momentum-dependent spectral weight. For a real tunneling process, the spectral weight contribution from certain momentum directions may dominate depending on the tunneling matrix element and the possible scattering from the surface layer \cite{HoffmanJE2011RPP}. This is why, for iron-based superconductors, only one gap can be detected on some surfaces \cite{KatoT2009,YinY2009PRL,FasanoY2010,HanaguriT2010Science,SongC2011}, whereas multiple gaps can be observed on other surfaces \cite{ShanL2011,ShanL2011prb,TeagueM2011}.

Beyond previous measurements \cite{ShanL2011,ShanL2011prb}, the most interesting finding here is a hump-dip structure located outside the coherence peaks, as indicated by black arrows in Fig.~\ref{fig:fig1}(b). In previous work \cite{ShanL2011prb}, our studied energy window was limited to $\pm$20 meV to avoid possible disturbance of higher voltage to the tunnel junctions. In that case, only the initial part of the hump-dip structure could be seen (see Figs. 2 and 4 in Ref.\cite{ShanL2011prb}). For superconductors with a clean superconducting DOS, one should observe such fine structures in the tunneling spectra. As shown in Fig.~\ref{fig:fig1}(b), the hump-dip structure is beyond weak-coupling BCS theory, but it is similar to the DOS feature of conventional strong-coupling superconductors. Figure~\ref{fig:fig1}(d) shows the second derivative of the tunneling current ($d^{2}I/dV^{2}$ vs. $V$) of Pb, where only the positive-bias part is plotted for clarity. By comparing to the BCS curve (in green), one can see two dips ascribed to the peaks in the phonon density of states $F(\omega)$ (phonon modes) following Eliashberg strong-coupling theory [Fig.~\ref{fig:fig1}(e)] \cite{McMillanW1969,WolfE1985}. High-bias features in tunneling spectra were also observed in copper oxide superconductors \cite{ZasadzinskiJF2001PRL, NiestemskiFC2007}, which were ascribed to a strong electronic coupling to spin resonance modes both by qualitative analyses \cite{ ZasadzinskiJF2001PRL,NiestemskiFC2007,ZhaoJ2011} and quantitative Eliashberg calculations \cite{ZasadzinskiJF2006PRL,AhmadiO2011PRL}. The $d^{2}I/dV^{2}$ vs. $V$ curve obtained for Ba$_{0.6}$K$_{0.4}$Fe$_{2}$As$_{2}$ also shows a distinct dip at the bias voltage (and a peak at corresponding negative bias) well above the gap edge as shown in Fig.~\ref{fig:fig1}(c). From Figs.~\ref{fig:fig1}(a)-(c), we see that this high-bias feature is weaker at negative bias than that at positive bias. In analogy to that of Pb and copper oxides superconductors, the dip shown in Fig.~\ref{fig:fig1}(c) in Ba$_{0.6}$K$_{0.4}$Fe$_{2}$As$_{2}$ may be related to some bosonic modes.

\begin{figure}
\includegraphics[scale=1.3]{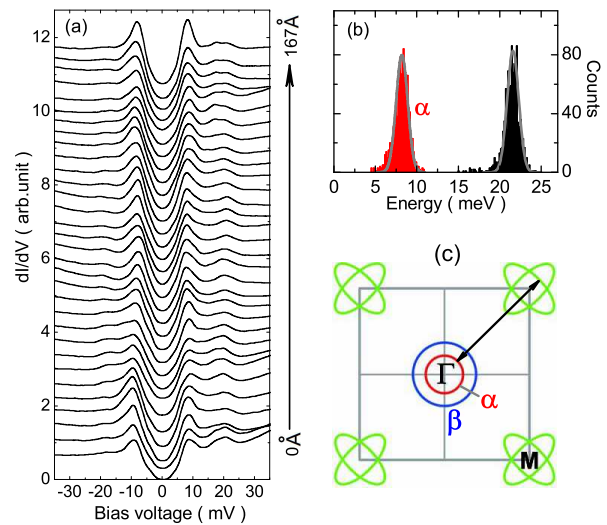}
\caption{\label{fig:fig2} (Color online) (a) Spatially-resolved tunneling spectra recorded along a line of 16.7 nm on the cleaved surface of a single crystal of Ba$_{0.6}$K$_{0.4}$Fe$_{2}$As$_{2}$. (b) A histogram of the occurrences of the larger gap $\bigtriangleup$ (red) and the dip $E_{dip}$ (black), which were observed in the same region mentioned in Fig.~\ref{fig:fig1}. (c) Schematic diagram of the Fermi surface pockets of Ba$_{0.6}$K$_{0.4}$Fe$_{2}$As$_{2}$ ($\alpha$-hole, $\beta$-hole pockets centered at $\Gamma$ point and electron pocket around M point). The black arrow indicates the nesting vector between the $\alpha$-hole pocket and the electron pocket.}
\end{figure}

Figure~\ref{fig:fig2}(a) shows a series of tunneling spectra recorded along a line with an interval of 5.4 {\AA}, indicating more homogeneous electronic states than those of copper oxide superconductors \cite{Fischer2007review}. We present in Fig.~\ref{fig:fig2}(b) the statistics of the larger superconducting gap ($\Delta$) around 8.4 meV mentioned above and the dip energy ($E_{dip}$) determined from a high-resolution $d^{2}I/dV^{2}$ map of the 16.7 nm $\times$ 8.4 nm region. In the superconducting state of a strong-coupling superconductor, a bosonic mode will appear in tunneling spectra at an energy offset by the gap, i.e., $\Omega=E_{dip}-\Delta$ in which $\Omega$ is the mode energy. In this way, we found a mode energy of $\Omega\approx 14$ meV connecting the larger gap around 8.4 meV mentioned above and the dip indicated in Fig.~\ref{fig:fig1}(c). For all studied regions, $\Omega$ was approximately 13-14 meV, which is very close to the energy of the neutron spin resonance mode of $\omega\approx 14$ meV as measured by neutron scattering both in polycrystalline Ba$_{0.6}$K$_{0.4}$Fe$_{2}$As$_{2}$ \cite{ChristiansonA2008} and in single crystalline Ba$_{0.67}$K$_{0.33}$Fe$_{2}$As$_{2}$ \cite{ZhangCL2011}. On the other hand, compared with angle resolved photoemission (ARPES) data \cite{DingH2008EPL,ZhaoL2008CPL}, the gap around 8.4 meV observed here stems mostly from the inner hole pocket in momentum space as indicated by $\alpha$ in Fig.~\ref{fig:fig2}(c). In fact, ARPES data show that the electron Fermi surface centered at the $M$ point also has a gap with a similar magnitude to that of the $\alpha$ pocket. However, because the tunneling rate along the $z$-axis is suppressed with increasing in-plane momentum \cite{TersoffJ1983}, the STS measurements presented here are more sensitive to states near the $\Gamma$ point. We note that the neutron spin resonance was observed at the nesting vector between the $\alpha$ pocket and the electron pocket, and the mode energy derived from the tunneling data is also related to the nested Fermi pockets. This gives further evidence that the mode observed here may originate from the neutron spin resonance mode, similar to the case of copper oxide superconductors \cite{ZasadzinskiJF2001PRL,ZasadzinskiJF2006PRL,AhmadiO2011PRL, NiestemskiFC2007,ZhaoJ2011}.

\begin{figure}
\includegraphics[scale=1.4]{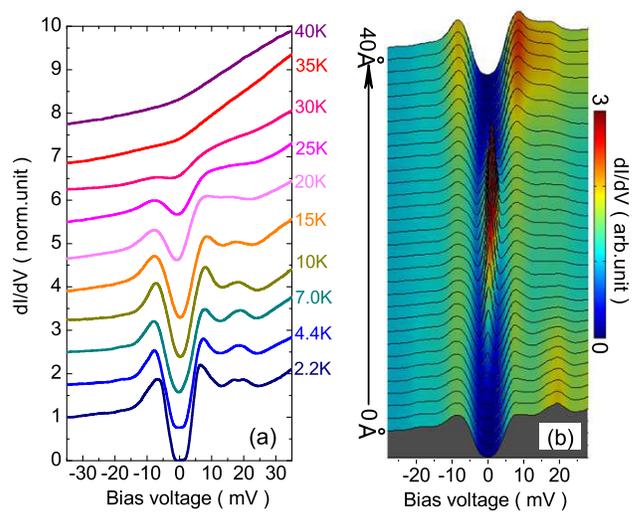}
\caption{\label{fig:fig3} (Color online) (a) Temperature dependence of averaged tunneling spectra. (b) Spatial dependence of tunneling spectra taken along a line cutting through a local impurity, indicating the close relationship between the strength of the hump-dip structure and the superconducting order parameter.}
\end{figure}

For a weak-coupling BCS superconductor, the pairing strength ($\bigtriangleup$) is assumed to be a real constant. For a strong-coupling superconductor, a complex and strongly energy-dependent gap function ($\bigtriangleup(E)$) is induced by a realistic electron-boson interaction, which is responsible for the fine features of the bosonic excitations in the tunneling spectra. To test the relationship between the bosonic features and the superconducting order parameter from the STS data, we studied the temperature and local impurity dependencies of the tunneling spectra. As shown in Fig.~\ref{fig:fig3}(a), the bosonic-excitation feature in the spectra became weak with increasing temperature and vanished near $T_c$. Furthermore, local impurities could also smear the feature as seen in the spatially-resolved STS measurements. Figure~\ref{fig:fig3}(b) shows the spectra measured along the trajectory of a 4-nm line crossing a single impurity. It is obvious that when the impurity states (behaving as a low-energy DOS peak) appear and hence the superconducting order is suppressed, the bosonic-excitation feature fades out. These data provide strong evidence that the bosonic mode is related to superconductivity.

Although the observed bosonic mode is consistent with the spin excitations in Ba$_{0.6}$K$_{0.4}$Fe$_{2}$As$_{2}$, one must also consider phonon contributions. Experimentally, the phonon density of states of BaFe$_2$As$_2$ measured by neutron scattering show a clear peak near 12 meV that has been identified as mostly arising from Ba vibrations \cite{MittalR2008PRB,ZbiriM2009PRB}. Upon K-doping to Ba$_{0.73}$K$_{0.27}$Fe$_2$As$_2$, the phonon modes near $E=10-15$ meV soften and broaden, but do not display any anomaly across the superconducting transition temperature $T_c$ \cite{LeeCH2010JPSJ}. The bosonic mode observed in our STS study has a close relationship with superconductivity and thus could not be the Ba-phonon modes that are not related to superconductivity. Moreover, the phonon modes related to the Fe-As layer have an energy scale of approximately 32 meV \cite{MittalR2008PRB,ZbiriM2009PRB}, which is much higher than the energy of the bosonic mode observed here.

\begin{figure}[]
\includegraphics[scale=1.5]{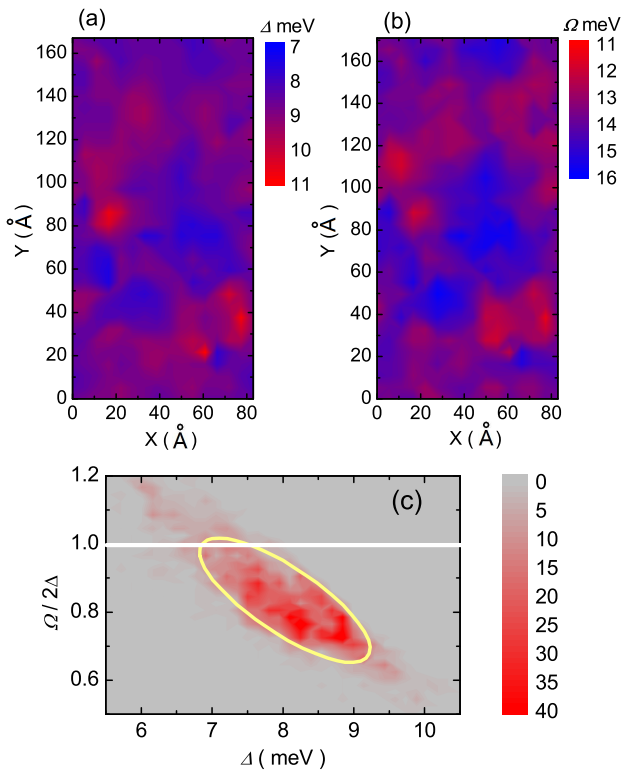}
\caption{\label{fig:fig4} (Color online) (a) Image of the superconducting gap values, or gap map ($\Delta(r)$). (b) Image of the mode energy $\Omega(r)=E_{dip}(r)-\Delta(r)$. Both (a) and (b) were obtained in the same region as mentioned in Fig.~\ref{fig:fig1} and Fig.~\ref{fig:fig2}. (c) Histogram of the occurrences of $\Omega$ plotted as a local ratio $\Omega(r)/2\Delta(r)$ against $\Delta(r)$. }
\end{figure}

Figures~\ref{fig:fig4}(a) and (b) show spatial distributions of the superconducting gap ($\Delta$) and spin-excitation mode energy ($\Omega$), respectively (the figures are plotted in the same way as that in the reference \cite{LeeJ2006Nature}). The similar patterns of the $\Delta$ map and $\Omega$ map indicate that $\Omega$ is locally anticorrelated with $\Delta$ (note that the color bar is reversed in Fig.~\ref{fig:fig4}(b)). This phenomenon is universal for all the regions observed here. The strong local anticorrelation suggests that the bosonic mode observed here is an electronic excitation with an intrinsic origin rather than a phonon mode or other extrinsic inelastic excitation unrelated to superconductivity.

It should be emphasized that the anticorrelation is local with a characteristic length of less than 2 nm and does not indicate an anticorrelation between bulk superconductivity ($T_c$ or bulk-averaged superconducting gap) and bulk-averaged spin-excitation energy. Such local anticorrelation is not accidental and has been observed in electron-doped copper oxide superconductors \cite{NiestemskiFC2007,ZhaoJ2011}. In addition, in these electron-doped cuprates, direct correlation between bulk superconductivity and bulk-averaged spin-excitation energy was observed simultaneously \cite{ZhaoJ2011}. This is consistent with the data from break junction tunneling obtained on hole-doped copper oxide superconductors Bi$_2$Sr$_2$CaCu$_2$O$_{8+\delta}$ (Bi2212) \cite{ZasadzinskiJF2001PRL}, in which a high-bias feature in the tunneling spectrum was also ascribed to the resonance spin excitation \cite{ZasadzinskiJF2001PRL}. Furthermore, a quantitative Eliashberg analysis of the Bi2212 data was done to extract the electron-boson spectral function, $\alpha^2F(\omega)$, suggesting a narrow boson spectrum as paring glue \cite{ZasadzinskiJF2006PRL,AhmadiO2011PRL}, consistent with the results of optical conductivity \cite{CarbotteJP2011RPP}. This encourages us to relate the observed single dip in the spectrum of $d^{2}I/dV^{2}$ vs. $V$ to a single, strong peak in $\alpha^2F(\omega)$. More interestingly, the ratio of $\Omega/2\Delta$ is always less than 1 for both hole-doped copper oxides and their electron-doped counterparts \cite{ZasadzinskiJF2001PRL,YuG2009NPhys,NiestemskiFC2007}, consistent with a magnetic exciton in the spin-fermion type models \cite{AbanovAr1999PRL,EschrigM2006AIP}. This is very similar to our data with $\Omega/2\Delta$ below 1 for a statistically significant fraction as shown in Fig.~\ref{fig:fig4}(c). At present, a quantitative Eliashberg analysis of the tunneling data for iron-based superconductors is difficult to carry out due to the multiband nature of these materials. Nevertheless, the observation of a strong coupling between the spin resonance and electron tunneling spectra in both the iron and copper based superconductors, in spite of their dramatically different parent compounds (semi-metals for iron based superconductors and Mott insulators for copper oxide superconductors), suggests that electron-spin coupling is an important general feature of these materials.

In summary, we have discovered a bosonic excitation in the spatially-resolved tunneling spectra of the hole-doped iron-based superconductor Ba$_{0.6}$K$_{0.4}$Fe$_{2}$As$_{2}$. This mode occurs near 14 meV and has a close relationship with the superconducting order parameter. Because the mode energy is very close to the neutron spin resonance, we argue that the resonance must arise from quasiparticle excitations between the inner $\alpha$-hole pocket and the electron pocket. This spin resonance mode is similar to that observed in copper oxide superconductors, supporting that the spin fluctuation mechanism may be a generic characteristic of high-Tc superconductors.

\begin{acknowledgments}
The authors are grateful for the kind help of Prof. Shuheng Pan during the STS measurements and thank Leland Harriger for critical reading of the paper. This work was supported by the National Natural Science Foundation of China (Project 11174349 and 51002180), the National Basic Research Program of China/973 programs (2011CBA00100, 2012CB821400, 2010CB833102, and 2010CB923002), and the Chinese Academy of Sciences. Work at UTK is supported in part by the US DOE, BES through DOE DE-FG02-05ER46202.
\end{acknowledgments}

\bibliography{BaKFeAs}

\end{document}